\documentclass[twocolumn]{aastex62}

\usepackage{newtxtext,newtxmath}
\usepackage[T1]{fontenc}
\usepackage{graphicx}	
\usepackage{amsmath}	

\usepackage{lineno}
\usepackage{xcolor}

\newcommand{\msun}{\textup{M}_\odot}
\newcommand{\lcdm}{\rm \Lambda CDM}

\begin{document}


\title[Numerics of SNe and Bursty Star Formation]{Bursty Star Formation in Dwarfs is Sensitive to Numerical Choices in Supernova Feedback Models}
\shorttitle{ Numerics of SNe Feedback }
\shortauthors{Zhang et. al.}

\author[0000-0002-7611-8377]{Eric Zhang}
\affiliation{Department of Physics and Astronomy, University of California, Riverside, CA 92507, USA}

\author[0000-0002-3790-720X]{Laura V. Sales}
\affiliation{Department of Physics and Astronomy, University of California, Riverside, CA 92507, USA}

\author[0000-0003-3816-7028]{Federico Marinacci}
\affiliation{Department of Physics \& Astronomy ``Augusto Righi'', University of Bologna, via Gobetti 93/2, 40129 Bologna, Italy}
\affiliation{INAF, Astrophysics and Space Science Observatory Bologna, Via P. Gobetti 93/3, 40129 Bologna, Italy}

\author[0000-0002-5653-0786]{Paul Torrey}
\affiliation{Department of Astronomy, University of Virginia, 530 McCormick Road, Charlottesville, VA 22903, USA}

\author[0000-0001-8593-7692]{Mark Vogelsberger}
\affiliation{Department of Physics, Kavli Institute for Astrophysics and Space Research, Massachusetts Institute of Technology, Cambridge, MA 02139, USA}

\author[0000-0001-5976-4599]{Volker Springel}
\affiliation{Max-Planck-Institut f\"{u}r Astrophysik, Karl-Schwarzschild-Stra\ss{}e 1, 85740 Garching bei M\"{u}nchen, Germany}

\author[0000-0002-1253-2763]{Hui Li}
\affiliation{Department of Astronomy, Tsinghua University, Haidian DS 100084, Beijing, China}

\author[0000-0003-3308-2420]{Rüdiger Pakmor}
\affiliation{Max-Planck-Institut f\"{u}r Astrophysik, Karl-Schwarzschild-Stra\ss{}e 1, 85740 Garching bei M\"{u}nchen, Germany}

\author[0000-0001-6179-7701]{Thales A. Gutcke}\thanks{NASA Hubble Fellow}
\affiliation{Institute for Astronomy, University of Hawaii, 2680 Woodlawn Drive, Honolulu, HI 96822, USA}

\begin{abstract}
Simulations of galaxy formation are mostly unable to resolve the energy-conserving phase of individual supernova events, having to resort to subgrid models to distribute the energy and momentum resulting from stellar feedback. However, the properties of these simulated galaxies, including the morphology, stellar mass formed and the burstiness of the star formation history, are highly sensitive to numerical choices adopted in these subgrid models. Using the {\small SMUGGLE} stellar feedback model, we carry out idealized simulations of a $M_{\rm vir} \sim 10^{10} \, \msun$ dwarf galaxy, a regime where most simulation codes predict significant burstiness in star formation, resulting in strong gas flows that lead to the formation of dark matter cores. We find that by varying only the directional distribution of momentum imparted from supernovae to the surrounding gas, while holding the total momentum per supernova constant, bursty star formation may be amplified or completely suppressed, and the total stellar mass formed can vary by as much as a factor of $\sim 3$. In particular, when momentum is primarily directed perpendicular to the gas disk, less bursty and lower overall star formation rates result, yielding less gas turbulence, more disky morphologies and a retention of cuspy dark matter density profiles. An improved understanding of the non-linear coupling of stellar feedback into inhomogeneous gaseous media is thus needed to make robust predictions for stellar morphologies and dark matter core formation in dwarfs independent of uncertain numerical choices in the baryonic treatment.
\end{abstract}


\section{Introduction}\label{sec:intro}

Galaxy formation at the scale of dwarfs ($M_{*} \sim 10^6 - 10^9 \, \msun$) is extremely inefficient, turning only a small fraction of the available gas into stars \citep[$\sim 10 \%$, e.g.,][]{Conroy09,Moster13,Behroozi19}. One of the most important processes affecting the formation of stars in dwarf galaxies are supernovae (SNe), which deposit a significant amount of energy into the interstellar medium (ISM) \citep[e.g.,][]{Larson74,Navarro93,Agertz13Energy}. This energy deposition is instrumental in regulating the rate of star formation in galaxies of all masses to the observed levels \citep[e.g.,][]{Agertz10,Guedes11,Hopkins11,Aumer13,GattoWalch16}. On the scale of dwarfs, supernova energy coupled efficiently to the ISM may also be responsible for generating fluctuating gravitational potentials that result in the  redistribution of dark matter from the center to the outer regions of the galaxy and the formation of constant density cores \citep[e.g.,][]{Navarro96,PontzenGovernato12,DiCintio14,Dutton16,ElBadry16,Read16,Tollet2016,Freundlich20,Jahn23,Azartash-Namin2024}.

In practice, a SNe event can be modeled as a point injection of a large amount of thermal energy, which expands as a shock into its environment. The first stage of this expansion is energy-conserving, as the thermal pressure inside the shock drives the bubble's expansion according to the Sedov-Taylor solution \citep[e.g.,][]{Taylor50,Sedov59}. Once the shock slows down and the driving pressure becomes comparable to the ambient pressure, the bubble enters the snow-plow or momentum-conserving phase, where the swept up matter continues to move without being subject to external forces.

However, it is very difficult for simulations to accurately capture the way in which SNe explosions expand into the ISM \citep[e.g,][]{Scannapieco12}. Firstly, the Sedov-Taylor solution, and related supernova feedback models \citep[e.g.,][]{Chevalier74,Cioffi88,DraineWoods91,Thornton98}, invoke a number of simplifying assumptions -- in particular, it is derived based on a homogeneous local ISM at rest. However, neither of these assumptions are necessarily true for SNe in galaxies; the local ISM may be inhomogeneous on a wide range of scales, including scales comparable to the size of the blast wave itself. It will also often not be at rest, and in many cases we should expect the gas to be moving away from active star-forming areas, due to previous SNe events or the action of stellar winds. Furthermore, the resolution of simulations is often insufficient to explicitly capture the relevant physics of the SNe blastwave. The time resolution needed to resolve the radiative cooling of very hot gas after a point injection of thermal energy is difficult to reach in galaxy formation simulations, leading to what is known as the `overcooling problem' \citep[e.g.,][]{Katz92,SmithSijackiShen18,Vogelsberger20}, in which, due to purely numerical errors, the thermal energy injected by supernovae dissipates too quickly to expand as a shock into the ISM. Often, the spatial resolution is also too coarse to individually resolve the energy-conserving and momentum-conserving phases of the expanding shock. Typical cooling radii, or size of the bubble at the onset of the momentum-conserving phase, span from $\sim$ pc-scale to a few hundred parsecs, depending on the local conditions of the ISM. This is often smaller than the spatial resolution of galaxy-scale numerical simulations, leaving the energy-conserving phase unresolved.

As such, SNe feedback is often implemented as a sub-grid process. Methods include modified thermal injection schemes \citep[e.g., ][]{Murante10,DallaVecchia12,Chaikin23Cosmo}, directly injecting momentum into the ISM \citep[e.g.,][]{Navarro93,Mihos94,Vogelsberger13,Hopkins18SNFeed}, temporarily disabling radiative cooling after SNe events \citep[e.g.,][]{Stinson06, Agertz10, Teyssier13}, modeling clusters of multiple SNe collectively as an expanding superbubble \citep[e.g.,][]{Keller14,ElBadry19SB}, or resorting to smoothly representing the ISM by means of an effective equation of state \citep{Springel2003}. Several current subgrid SNe models based on direct moment injection have begun to explicitly compensate for resolution limitations by injecting the terminal momentum of the SNe ejecta after the unresolved Sedov-Taylor phase, accounting in the calculation for various dependencies including, local gas density and metallicity \citep[e.g.,][]{KimOstriker15, Pittard19, Hopkins18SNFeed, SMUGGLE, Karpov20}. Notably, {\small FIRE-3} \citep[][]{FIRE3} adjusts the momentum injected also for instances when the medium is not at rest, in an attempt for a more physically-motivated model. In general, higher order considerations and more detailed sub-grid treatments may be advisable \citep{Hopkins2024}.

Recently, some simulations have reached the ultra-high resolution needed to model SNe via direct injection of thermal energy \citep[e.g.,][]{Emerick19,Hu19,Lahen20,Lyra2021,Deng24RIGEL}, and can thus explicitly resolve the expansion of SNe bubbles into realistic, inhomogeneous ISM structures. At these resolutions, the momentum injection into the ISM will naturally arise from the thermal expansion, and will thus not rely on arbitrary numerical choices. However, these resolutions are not yet achievable in large-scale simulations. The reality of SNe injection in galaxy formation simulations is, likely, substantially more complex than assumed in the prescriptions used, mostly due to the (unresolved) inhomogeneities in the ISM of galaxies.

For example, even a modestly low density of $n \sim 0.01$ particles per ${\rm cm}^3$ corresponds to a cooling radius of about $200$ pc, which is often comparable to the size of superbubbles formed by clustered SNe feedback \citep[e.g.,][]{Li24SBs}. Studies using small-scale simulations of a more realistic inhomogeneous ISM \citep[e.g.,][]{Martizzi15} have shown departures in the evolution of the SNe remnants from the traditionally assumed sub-grid models that rest on assumptions of spherically symmetric expansion into an homogeneous media. The propagation speed and momentum coupling of the SNe energy vary along low- and high-density channels within the structured ISM, contributing to the departure from idealized solutions in homogeneous media. On a macroscopic level, for stratified media (such as the disk of galaxies) this can result in preferential coupling of the energy/momentum for different directions, which might not be properly taken into account when enforcing a symmetric energy/momentum deposition in subgrid models. 

It is necessary to produce accurate subgrid models of SNe feedback since many of the predicted properties of simulated galaxies may show variations depending on the details of the baryonic treatment and, in particular, the stellar feedback implementation \citep{Sales10, Scannapieco12, Chaikin22Numeric}. This is particularly true in the regime of dwarf galaxies, where the relatively shallower gravitational potential makes the system more responsive to the energy and momentum input from feedback. There is consensus in the field that stellar feedback, and mostly SNe, are capable of regulating star formation in dwarf galaxies to observed levels, by disrupting cold, dense gas clumps that would form far too many stars compared to observations if not regulated by feedback \citep[e.g., see comparison of several codes in ][]{Sales22}. However, there is less agreement on the level of ``burstiness" that is reasonable to expect in the regime of dwarfs, and whether theoretical models compare well with observational constraints \citep[e.g., ][]{Sparre17,Patel2018,Emami2019,Emami2021}. 

If star formation is indeed as bursty as predicted by some of the state-of-the-art numerical simulations, this carries important consequences for the theoretical predictions on the stellar sizes, morphologies and inner dark matter distribution in dwarf galaxies. If gas contributes sufficiently to the local density in the inner regions of halos, which may occur in the case of gas-rich dwarfs, bursty episodes of star formation can generate substantial gas flows that create fluctuating gravitational potentials in the inner regions of low mass halos combined with powerful outflows \citep[e.g.,][]{MCS2021}. These quick (i.e., on sub-dynamical timescales) fluctuations in the gravitational potential of the galaxy result in a net gaining of energy for collisionless components like stars and dark matter, making them less tightly bound, and thus they tend to migrate outwards \citep{Navarro96,PontzenGovernato12,BenitezLlambay19}. Shallower distributions of stars and dark matter are thus expected in systems where star formation is bursty and stellar feedback is efficiently coupled to the surrounding ISM, including a transformation from cuspy dark matter halos into cored ones. Through this process, numerical details on the sub-grid modeling of star formation and stellar feedback have a direct impact on cosmological predictions and our understanding of CDM as plausible dark matter model. 

In this paper, we revisit the assumption of an isotropic deposition of SNe momentum for sub-grid models of feedback using the multi-phase {\small SMUGGLE} model \citep[e.g.,][]{SMUGGLE} applied in simulations of an isolated dwarf halo. In Sec.~\ref{sec:simulations} we discuss the simulation set-up and the methods used to control both: $i)$ directional anisotropy, and $ii)$ cell weighting scheme of the SNe feedback model. In Sections~\ref{sec:anisotropy} and \ref{sec:weighting} we present our main results on how varying these two assumptions may impact the large-scale dwarf properties, including their star formation histories, morphologies and inner dark matter distributions. In Section~\ref{sec:conclusion} we summarize our results.

\section{Description of the Simulations}\label{sec:simulations}

We use the moving-mesh hydrodynamics code {\small AREPO} \citep[][]{AREPO2010} to run simulations of an isolated $M_{\rm vir} \sim 10^{10} \, \msun$, $M_{*} \sim 10^{8} \, \msun$ dwarf galaxy where virial quantities are defined within the radius that encloses an average density of $200$ times the critical density of the Universe. Our runs have a reference resolution with  $\sim 6000$ $\msun$ for the baryonic component (gas and stars) and $\sim 25000$ $\msun$ for the dark matter particles. A gravitational softening length of 32 pc is used for every particle. Stars form by transforming an eligible gas cell into a single star, which means that initially stars have the same mass than the parent gas cell. The mesh refinement scheme implemented in {\sc AREPO} ensures that the mass of any given gas cell is within a factor of 2 from this reference resolution. Hence, star particles are formed at around this particle mass; however, due to mass loss in stellar feedback processes, star particle masses may decrease below this value; for example, in our default runs we find that the average stellar particle mass falls to $5000 \pm 1500\; \msun$ by the end of the simulation (where the uncertainty corresponds to the r.m.s of the stellar particle mass distribution.) We test convergence of the results using a $\times 10$ better resolution run in the Appendix. 

The initial conditions consist of a gas-rich disk ($M_{\rm gas} \sim 5 \times 10^8 \, \msun$) in hydrostatic equilibrium with the dark matter halo, with an additional population of old disk stars ($M \sim 7 \times 10^7 \, \msun$) and old bulge stars ($M \sim 5 \times 10^6 \, \msun$). The initial dark matter halo is an NFW \citep[][]{NFW} profile with concentration $c=15$, the initial gas and stellar disks are in an exponential profile, and the initial stellar bulge is a \citet{HernquistProfile} profile. This set up is similar to those presented in \citet{Hopkins11, Burger22a, Burger22b, Jahn23}. The galaxy lies at the center of a 3D box with 200 kpc on a side. All simulations are run for at least 2 Gyr, with a snapshot saved every 5 Myr. All galaxy-wide properties discussed within this work are calculated from all snapshots over the first 2 Gyr, unless stated otherwise.

We briefly summarize the {\small SMUGGLE} stellar feedback model as per \citet{SMUGGLE} (hereafter M+19). It includes various processes of star formation, stellar feedback and the heating and cooling of gas. Star formation is primarily modulated by two numerical parameters: the star formation efficiency $\epsilon_{\rm SF}$, which regulates the rate at which a given gas cells form stars, and a density threshold $\rho_{\rm th}$, which gas cells are required to exceed in order to form stars. Chief among the relevant feedback mechanisms, and the one we will focus on in this paper, is the feedback done by supernovae.

The main parameters involved in the star-formation and feedback modeling are summarized in Table \ref{tab:parameters}. Our findings are robust to a range of numerical choices, which is presented in Appendix~\ref{app:numerics}.

\begin{table}
\centering
\begin{tabular}{lr}
    \hline
    Parameter & Value \\
    \hline
    Halo Virial Mass & $1 \times 10^{10} \, \msun$ \\
    Disk Stellar Mass (Initial) & $7 \times 10^{7} \, \msun$ \\
    Bulge Stellar Mass (Initial) & $5 \times 10^{6} \, \msun$ \\
    Gas Mass (Initial) & $5 \times 10^{8} \, \msun$ \\
    \hline
    Target Baryonic Mass & 6000 $\msun$ \\
    Dark Matter Particle Mass & 25000 $\msun$ \\
    Softening Rad. (All Particles) & 32 ${\rm pc}$ \\
    \hline
    SF Efficiency & 0.01 \\
    SF Dens. Threshold & 100 $m_{\rm H}/{\rm cm^3}$ \\
    \hline
\end{tabular}
\caption{List of the numerical parameters and initial conditions used in the simulations.}
\label{tab:parameters}
\end{table}

\subsection{Star Formation Prescription}

To model star formation, gas cell are converted into star particles, probabilistically, at a rate consistent with
\begin{equation}
    \dot{M}_{\rm SF} = \epsilon_{\rm SF} \frac{M_{\rm cell}}{t_{\rm dyn}},
\end{equation}
where $\epsilon_{\rm SF}$ is the star formation efficiency, which is a tunable dimensionless factor which we set to 0.01 \citep[M+19, in accordance with][]{KrumholzTan07}, $M_{\rm cell}$ is the mass of the gas cell and $t_{\rm dyn}$ is the dynamical timescale of the gas within the cell, given by:
\begin{equation}
    t_{\rm dyn} = \sqrt{\frac{3 \pi}{32 G \rho_{\rm cell}}},
\end{equation}
where $G$ is the gravitational constant and $\rho_{\rm cell}$ is the density of the gas cell.

Additionally, gas cells must be sufficiently dense in order to form stars; this is enforced by allowing only gas cells with a density $\rho_{\rm cell} > \rho_{\rm th}$ to become a star particle. In the main body of this work, we set $\rho_{\rm th}$ to be $100$ particles per ${\rm cm}^3$. The {\small SMUGGLE} model also requires that the gas is locally gravitationally bound in order to form stars; this is modeled by a virial parameter criterion for each gas cell, described in M+19; only gas cells satisfying this criterion are eligible to become a star particle.

At the time of formation, star particles keep the mass, position, and velocity of their progenitor gas cells. Each star particle represents a stellar population, consistent with a \citet{ChabrierIMF} initial mass function, which evolves coevally according to the model described in M+19.\footnote{For mass resolutions within this work, the mass of star particles are sufficiently high that each particle can be treated as a stellar population that samples well the IMF.}

\subsection{Supernova Feedback} \label{subsec:snfeed}

SNe events occur according to the stellar population evolution model described in M+19, in which Type II or Ia supernovae occur in a star particle according to the ages of OB and white dwarf stars within a given stellar population.  SNe are resolved discretely; typically, at the resolution of our simulations, no more than 1 SN occurs within a star particle at any given timestep.

At each supernova event, a nominal energy of
\begin{equation}
    E_{\rm tot} = n_{\rm SN} E_{51} \,,
\end{equation}
is released by a star particle into the neighboring gas particles where $n_{\rm SN}$ is the number of supernovae occurring within the numerical event, and $E_{51} = 1.0 \times 10^{51}$ ${\rm erg}$. This energy is entirely converted into a scalar momentum, via
\begin{equation}
    p_{\rm tot} = \sqrt{ 2m_{\rm tot} E_{\rm tot} },
\end{equation}
where $m_{\rm tot}$ is the total mass of the supernova ejecta.
This linear momentum is then partitioned among the nearest $N_{\rm ngb} = 32$ gas cells, according to a weight factor $w_i$ for each neighboring cell, indexed by $i$. Then, the (nominal) linear momentum injected into each neighbor gas cell is
\begin{equation}
    \Tilde{p}_i = \frac{ p_{\rm tot} w_i }{ \sum_i w_i },
\end{equation}
so that, in principle, the sum of all linear momenta entering the gas cells is $p_{\rm tot}$. However, a number of numerical `post-processing' procedures are applied before imparting the momentum to a gas cell.
Firstly, due to the unresolved nature of the energy-conserving phase of the SN blast wave in these simulations, the linear momentum injected into a gas cell $i$ by a star particle is boosted\footnote{Due to the typical value of $p_{\rm term}$ given in Eq.~(\ref{eq:pterm}), this value is in practice always greater than 1.} \citep[in a manner consistent with][]{Hopkins18SNFeed} by a factor
\begin{equation}\label{eq:boostfac}
    \beta = \min \left(
        \sqrt{ 1 + \frac{m_i}{\delta m_i} },
        \frac{ p_{\rm term} }{ p_{\rm tot} }
    \right),
\end{equation}
where $m_i$ is the mass of the target gas cell, $\delta m_i = m_{\rm tot} w_i / \sum_i w_i$ is the mass of the ejecta entering the gas cell, and $p_{\rm term}$ is the terminal momentum of the SNe shock, which is \citep[in accordance with][and used by M+19]{Cioffi88}, given by
\begin{equation}\label{eq:pterm}
    p_{\rm term} =
    4.8 \times 10^5 \, \msun \, {\rm km} \, {\rm s}^{-1}
    \cdot \bar{E}^{ 13/14 }
    \bar{n}^{ -1/7 }
    f(Z) \, ,
\end{equation}
where $\bar{E}$ is the total energy released in units of $10^{51}$ erg, $\bar{n}$ is the local density estimate in units of particles per ${\rm cm}^3$, and
\begin{equation}
    f(Z) = \min \left[ 
    \left(\frac{Z}{Z_{\odot}} \right)^{-0.14} , 2 
    \right]^{3/2} \, ,
\end{equation}
where $Z$ is the metallicity, which for simplicity is set to the solar value ($Z_\odot$). The inclusion of the minimum function in Eq.~(\ref{eq:boostfac}) ensures that even when the SN momentum is boosted, the total momentum injected does not exceed the maximum possible momentum injected. The actual linear momentum injected into a gas cell is then
\begin{equation}
    p_i = \beta \Tilde{p}_i \, .
\end{equation}
Furthermore, the nearest $N_{\rm ngb}$ neighboring gas cells are sometimes too far away from the star particle for momentum injection to be considered physical. To prevent star particles from being able to affect far-away gas cells, a `superbubble limiter radius' of $r_{\rm SB} = 1024 \, {\rm pc}$ is imposed on all star particles, preventing momentum from being injected into gas cells further than this radius.\footnote{M+19 verifies that varying this radius within a large range of values does not change the results.}

The default choice for the weight is the solid angle of opening $\Omega_i$, defined by
\begin{equation}\label{eq:omega}
    \Omega_i = 2 \pi \left\{
    1 - \frac{1}{ 
    \left[ 1 + A_i / (\pi | {\bf r}_i - {\bf r}_{*} |^2) \right]^{1/2} 
    }
    \right\} \ ,
\end{equation}
where $A_i$ is the cell boundary area. Note that this weighting scheme follows that in other models, such as {\small FIRE} \citep[see Eq. 2 in][]{Hopkins18SNFeed}. However, other choices for the cell weights are possible, such as the kernel-weighted cell volume $w_i = W_i V_i$ and the kernel-weighted cell mass $w_i = W_i m_i$ (where $V_i$ is the volume of the target gas cell, and $W_i = W( |{\bf r}_i - {\bf r}_{*}|, h)$ is the standard cubic spline SPH kernel value).

\begin{figure}
\includegraphics[width=0.98\columnwidth]{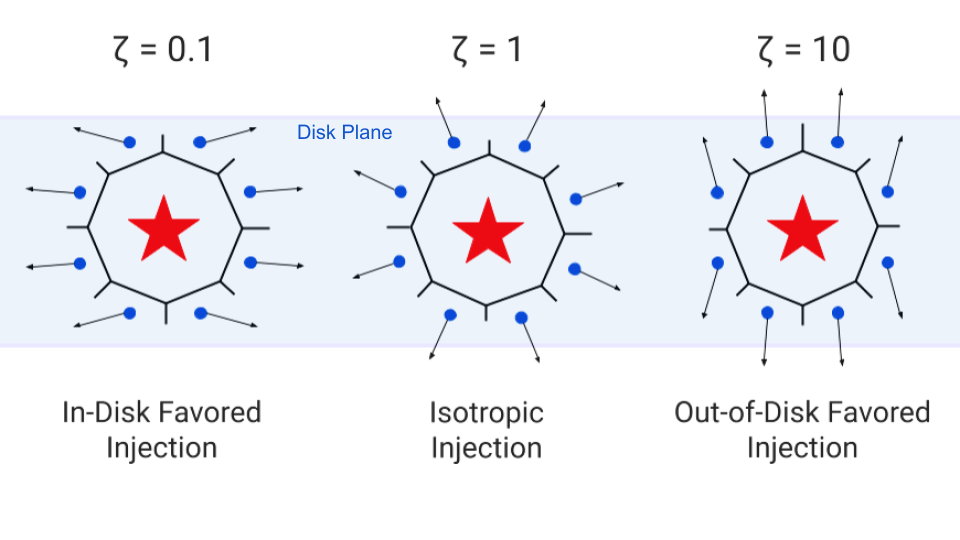}
    \caption{Picture of the anisotropic SNe momentum injection scheme, as parametrized by the nominal anisotropy factor $\zeta$. The star particle undergoing a SNe event is shown in red, and the surrounding gas cells are shown as blue circles. The arrows indicate the magnitude and direction of the momentum injection. A low $\zeta$ value injects momentum primarily within the disk plane, and a high $\zeta$ injects momentum primarily out of the disk plane. $\zeta=1$ implies isotropic injection. Note that only the direction of injected momenta varies with $\zeta$; the magnitude does not.}
    \label{fig:injection_cartoon}
\end{figure}

The weight factor $w_i$ controls the magnitude of momentum that each gas cell receives. The direction of this momentum injection can then independently be determined by the nominal anisotropy factor, $\zeta$. In particular, we control the vector momentum injected into each gas cell according to an anisotropic direction vector ${\bf k}$, such that the momentum ${\bf \delta p_i}$ delivered to the gas particle is
\begin{equation}\label{eq:zeta_def1}
    {\bf \delta p_i} = \frac{ p_i {\bf k} }{ \left| {\bf k} \right| } \ ,
\end{equation}
where the anisotropic direction vector ${\bf k}$ is defined by
\begin{equation}\label{eq:zeta_def2}
    {\bf k} = \left[ x_i - x_{*} \, , y_i - y_{*} \, , \zeta(z_i - z_{*}) \right] \ ,
\end{equation}
where $(x_i,y_i,z_i)$ is the position of the target gas particle, and $(x_{*},y_{*},z_{*})$ is the position of the star particle doing the feedback. According to this formula, the relative amount of momentum being injected in the $z$-direction is controlled by the factor $\zeta.$ We show a picture of this injection scheme in Fig. \ref{fig:injection_cartoon}.

A higher $\zeta$ means more momentum is being injected in the $z$-direction, but the normalization factors ensure that the magnitude of momentum released into a particular gas cell is independent of $\zeta$. $\zeta = 1$ implies isotropic injection; that is, the momentum injected into the cell will always point radially away from the source star particle. $\zeta$ can only change the direction of the injected momentum; for any given cell, the magnitude of momentum injected into it is not affected by the choice of $\zeta$; it is only controlled by the cell weighting scheme $w_i$ and boost factor $\beta$.

\begin{figure}
\includegraphics[width=0.95\columnwidth]{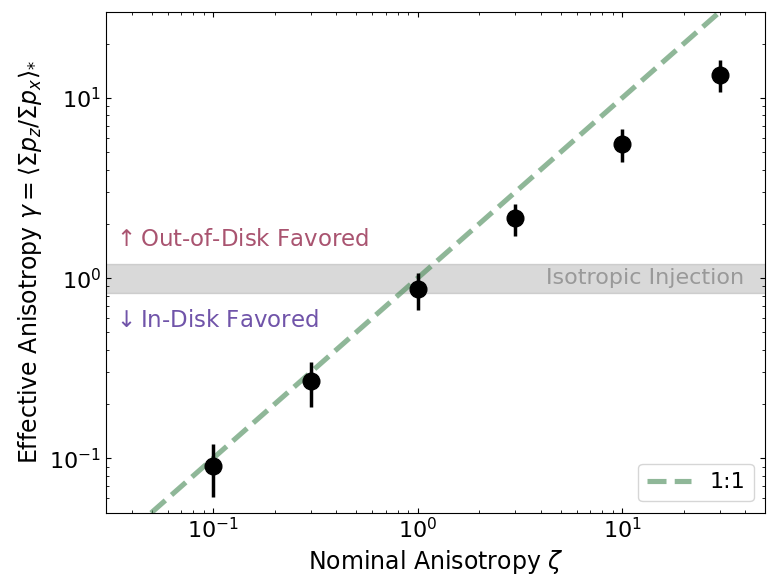}
    \caption{The effective anisotropy ($\gamma$, as defined in Eq. \ref{eq:gamma}), and its associated uncertainty, as a function of the nominal anisotropy ($\zeta$, as defined in Eqs. \ref{eq:zeta_def1}, \ref{eq:zeta_def2}). The nominal anisotropy controls the ratio of momentum injected in the polar ($z$) direction to the in-plane ($x,y$) directions, \textit{by every star particle, before `post-processing' effects are considered}. The effective anisotropy measures the ratio between the total momenta actually injected in these directions.}
    \label{fig:gammafac}
\end{figure}

As mentioned before, some post-processing calculations may affect the final momentum injected into the cell after the weight and direction has been defined. This occurs due to variations in the distribution of gas cells about star particles, which can affect the boost factor (Eq.~\ref{eq:boostfac}) and superbubble limiter. As a result, the nominal anisotropy $\zeta$ may be different from the effective anisotropy imparted to the individual cells. We quantify the true anisotropy in momentum injection by an effective anisotropy factor, $\gamma$, which we define by the ratio of the total momentum injected into gas by SNe along the $z$-axis (regardless of the positive or negative direction), to the total momentum injected along the $x$-axis\footnote{It would also be valid to use the $y$ axis for this definition, since it is also within the disk plane; however, we will choose to use the $x$ axis.},
\begin{equation}
    \gamma = 
    \left< 
    \frac{\sum_{\rm SNe} \sum_{i}  \delta p_{i,z}}{\sum_{\rm SNe} \sum_{i}  \delta p_{i,x}}
     \right>_{\rm stars}.
    \label{eq:gamma}
\end{equation}
In practice, this factor \textit{does not}  need to be equal to the nominal anisotropy factor $\zeta$. However, Fig. \ref{fig:gammafac} indicates that $\gamma$ does in fact scale well with $\zeta$, so that the degree of anisotropy is still reliably controlled by $\zeta$. Furthermore, $\gamma$ carries an associated uncertainty, which is the standard deviation of the quantity averaged over all star particles in Eq.~(\ref{eq:gamma}).

Thus, we explore the effect of anisotropy by varying $\zeta$, and we label our runs by the value of $\zeta$ we employ for that run. However, we will measure all properties in terms of the effective anisotropy, $\gamma$. Changes in the cell weighting scheme will also be discussed. Table~\ref{tab:allsims} shows a list of all runs covered in the main body of this paper, the value of $\zeta$ used, and the resulting value of $\gamma$ for the corresponding simulation.

\begin{table*}
\centering
\begin{tabular}{lcccr}
    \hline
    Simulation & Nominal SNe & Effective SNe & SNe Cell Weight & log10 $M_{*}$ \\
    Name & Anisotropy $\zeta$ & Anisotropy $\gamma$ & Scheme $w_i$ & Formed [$\msun$] \\
    \hline
    {\it zeta0.1} (In-Disk Favored) & 0.1 & 0.091 $\pm$ 0.030 & Solid Angle ($\Omega$) & 7.80 \\
    {\it zeta0.3} & 0.3 & 0.27 $\pm$ 0.08 & Solid Angle ($\Omega$) & 7.71 \\    
    {\it zeta1} (Isotropic) & 1.0 & 0.87 $\pm$ 0.20 & Solid Angle ($\Omega$) & 7.71 \\
    {\it zeta3} & 3.0 & 2.2 $\pm$ 0.4 & Solid Angle ($\Omega$) & 7.58 \\
    {\it zeta10} (Out-of-Disk Favored) & 10.0 & 5.6 $\pm$ 1.1 & Solid Angle ($\Omega$) & 7.38 \\
    {\it zeta30} & 30.0 & 13 $\pm$ 3 & Solid Angle ($\Omega$) & 7.32 \\
    \hline
    {\it zeta1-VolWgt} & 1.0 & 0.90 $\pm$ 0.23 & Kernel-Weighted Volume & 7.81 \\
    {\it zeta10-VolWgt} & 10.0 & 5.4 $\pm$ 1.4 & Kernel-Weighted Volume & 7.62 \\
    {\it zeta1-MassWgt} & 1.0 & 0.95 $\pm$ 0.30 & Kernel-Weighted Mass & 7.36 \\
    {\it zeta10-MassWgt} & 10.0 & 5.6 $\pm$ 1.9 & Kernel-Weighted Mass & 7.37 \\
    \hline
\end{tabular}
\caption{List of the simulations and the parameters used in each. Simulations are named by the value of \textit{zeta} ($\zeta$) and the cell weighting scheme used for that run, but we will quantify any physical properties by their dependence on the effective anisotropy $\gamma$, and its associated uncertainty. $M_{*}$ formed refers to the total stellar mass formed within the entire simulation volume, by the final simulation time of 2 Gyr.}
\label{tab:allsims}
\end{table*}

In theory, differing distributions of gas cells around star particles throughout the runs can lead to some variation in the \textit{total magnitude} of momentum being injected by a typical star particle between runs. In this case, we would be unable to conclude that differences in the dwarf properties are purely due to a change in the directional distribution of momentum. However, we do find that, for the \textit{same choice of cell weighting scheme}, $w_i$, varying the anisotropy factor does not change the typical momentum released and imparted to surrounding cells by a star particle\footnote{Since the momentum injected is measured on a per star-particle basis, it will depend on the simulation mass resolution as well. We do not consider this dependence in this work, since we use the same mass resolution for all simulations in the main body of this work.} to any significant degree. However, between \textit{different} cell weighting schemes, the total momentum injected can differ by more than half an order of magnitude. We examine the effect of the cell weighting scheme in more detail in Section~\ref{sec:weighting}.

\subsection{Other Physical Processes Involved In SMUGGLE}

In addition to SNe feedback, the {\small SMUGGLE} model includes other forms of stellar feedback, including OB and AGB stellar winds, radiation pressure, and photoionization of nearby gas. These modes of stellar feedback are energetically subdominant (M+19, Fig.~13), but can be important for pre-processing gas before SNe go off. Photoionization, in particular, is a crucial process in regulating star formation by disrupting star-forming gas clumps \citep[e.g.,][]{MCS2021}; it is enabled in our simulations, though we do not explicitly study its effects in this work. Also included are radiative heating and cooling mechanisms of the ISM gas, including cooling of H, He, and metals down to 10 K, self-shielding from ionizing UV radiation, cosmic ray heating, and photoelectric heating. The parameterization of these additional feedback processes do not change across different runs, so the only variation between the runs considered in this work are those of the SNe feedback.

\begin{figure*}[t]
\centering
\includegraphics[width=0.90\textwidth]{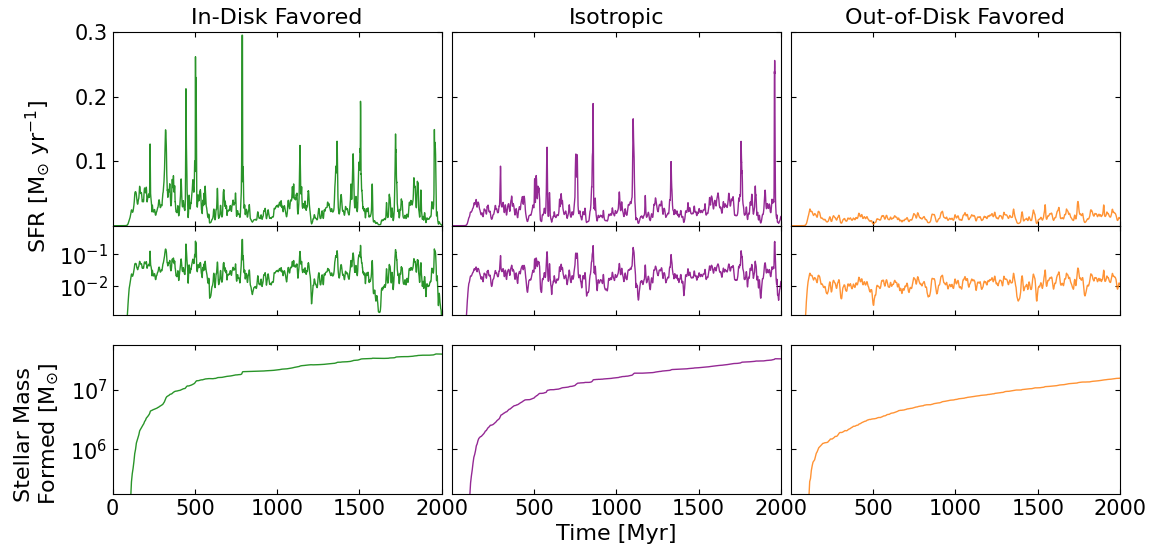}
    \caption{SF histories of the entire galaxy in the in-disk favored (green, left), isotropic (purple, center) and out-of-disk favored (orange, right) momentum injection runs. The top row shows the SFR in linear scale, the middle row shows the SFR in log scale, and the bottom row shows the cumulative stellar mass formed.} The burstiness of the SFR depends substantially on the injection model, with the in-disk favored and isotropic models resulting in more bursty histories and a greater stellar mass than the out-of-disk favored model.
    \label{fig:SFR_basic}
\end{figure*}

\section{Anisotropic Momentum Injection}\label{sec:anisotropy}

We explore the effects of varying the subgrid SNe momentum injection scheme. In this section, we focus on the effects of changing the directionality of the momentum injection. As introduced in Sec.~\ref{sec:simulations} and Eq.~(\ref{eq:gamma}), the anisotropy in SNe momentum injection is quantified by $\gamma$, where $\gamma=1$ indicates isotropic injection, and a $\gamma$ of greater (less) than 1 indicates momentum primarily being injected perpendicular to (within) the plane of the disk. We will focus on three runs in this section: \textit{zeta0.1} (in-disk favored injection), \textit{zeta1} (isotropic injection, default {\small SMUGGLE} choice), and \textit{zeta10} (out-of-disk favored injection). Where possible we will refer to specific runs by their descriptive indicators (i.e., `isotropic' for `\textit{zeta1}'). Each galaxy is followed for at least 2 Gyr.

\subsection{Star Formation History}

Of particular interest is the effect on the star formation (SF) history and its burstiness. A bursty star formation history can determine the mass, size and morphology of dwarfs, along with their dark matter distribution \citep[e.g.,][]{Stinson07,PontzenGovernato12,DiCintio14,GonzalesSamaniego14,ElBadry16,Sparre17}. We plot the star formation histories of the in-disk favored (\textit{zeta0.1}), isotropic (\textit{zeta1}), and out-of-disk favored (\textit{zeta10}) injection runs in Fig. \ref{fig:SFR_basic}.

Qualitatively, in-disk favored (green curve) and isotropic injection (purple curve) are associated with bursty star formation histories, as seen by the prevalence of sharp peaks in their SFR curves; contrarily, out-of-disk favored injection (orange curve) leads to a smoother star formation history.

\begin{figure}
\includegraphics[width=0.93\columnwidth]{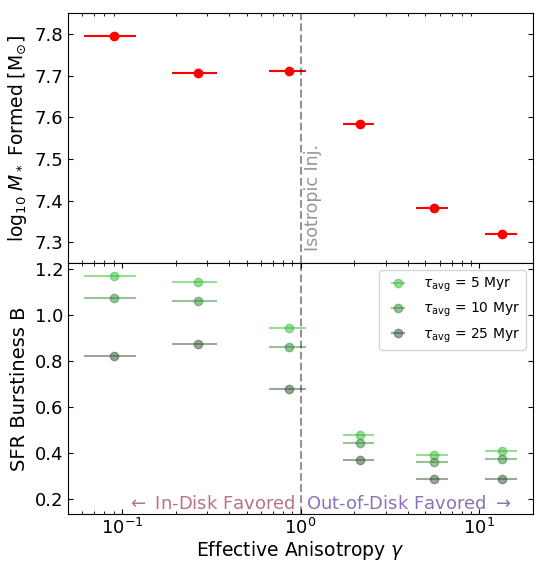}
    \caption{{\it Top:} Total stellar mass formed. Higher-$\gamma$ runs (i.e., runs with more momentum directed outside the disk), form a lower total stellar mass. The difference in the stellar mass formed can vary by up to half an order of magnitude, across all runs considered. {\it Bottom:} Burstiness parameter $B$ as defined by Eq.~(\ref{eq:burstiness}). Higher-$\gamma$ runs tend to be less bursty according to this parameter, over a wide range of SFR averaging timescales. The horizontal error bars are the uncertainties in $\gamma$, as discussed in Eq.~(\ref{eq:gamma}) and Table~\ref{tab:allsims}. }
    \label{fig:burstysfr}
\end{figure}

\begin{figure*}[t]
\centering
\includegraphics[width=0.95\textwidth]{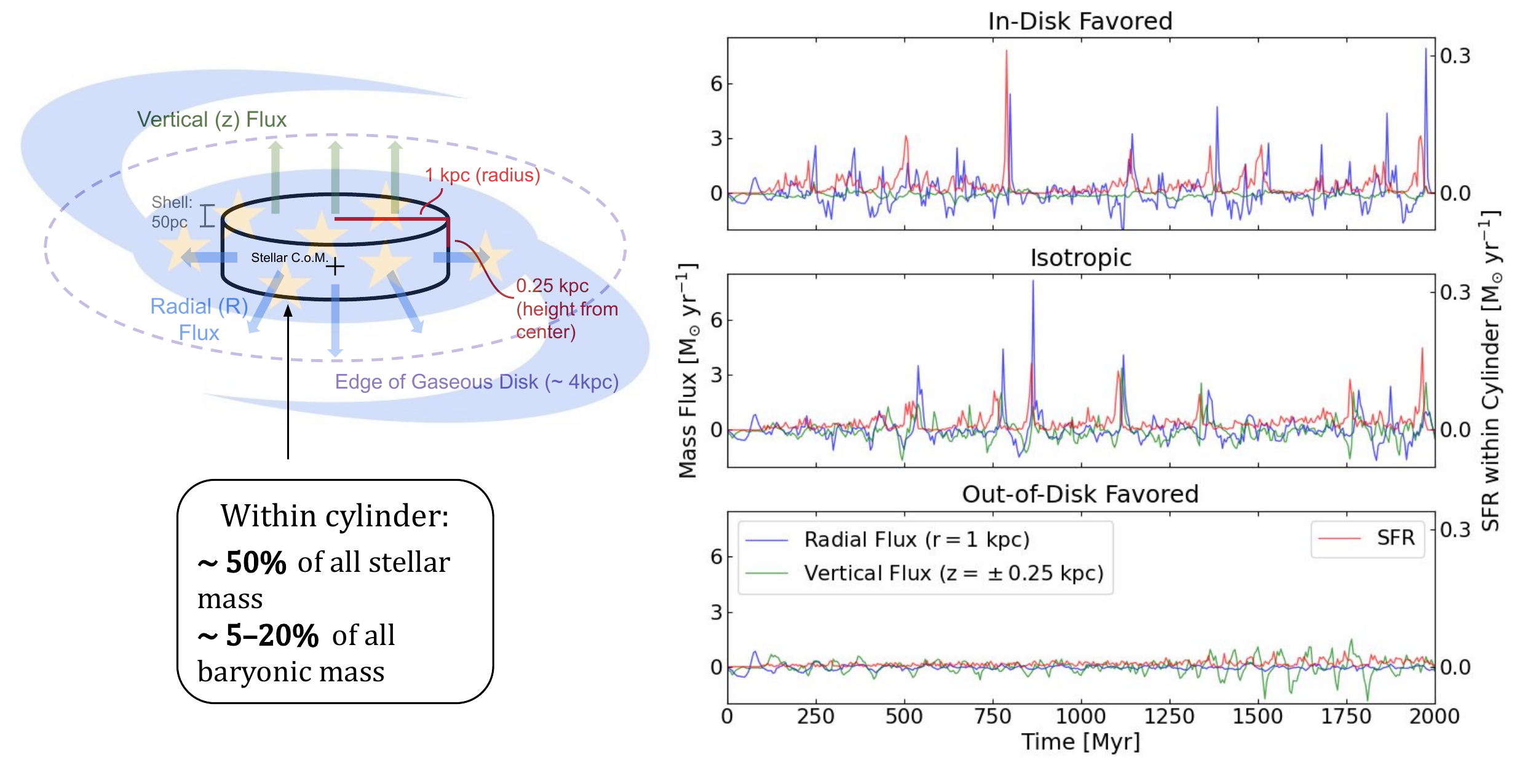}
    \caption{\textit{Left:} Diagram (not to scale) of the cylindrical region of interest, around which mass flux of gas are calculated. \textit{Right:} Mass flow history across the boundary of the cylinder of interest, of the in-disk favored (\textit{zeta0.1, top panel}), isotropic (\textit{zeta1, middle panel}) and out-of-disk favored (\textit{zeta10, bottom panel}) runs, and the associated SF history (red) in the interior of the cylinder. Radial gas flows (Eq. \ref{eq:mdotR}) are shown in blue, and vertical flows (Eq. \ref{eq:mdotZ}) are shown in green. Notably, in the isotropic run, starburst episodes (red peaks) are preceded by a inward bulk gas motion (blue and green dips) followed by peaks of outward motion (blue and green peaks) within the disk, consistent with a more turbulent gas disk. In the in-disk favored run, the vertical gas cycles are suppressed, and in the out-of-disk favored run, the radial gas cycles are suppressed. However, only in the out-of-disk favored run is bursty star formation and the overall mass flows suppressed, indicating that radial gas flows are more effective at driving turbulence.}
    \label{fig:gasmovement}
\end{figure*}

We quantify the burstiness of a run by the parameter
\begin{equation}\label{eq:burstiness}
    B = \frac{ \sigma \left( \langle
        {\rm SFR}
    \rangle_{\tau_{\rm avg}} (t) \right)}
    { \langle
        {\rm SFR}
    \rangle},
\end{equation}
where $B$ stands for the burstiness, $\langle{\rm SFR} \rangle_{\tau_{\rm avg}} (t)$ denotes the rolling average of the SFR over an interval $\tau_{\rm avg}$, $\sigma$ denotes the standard deviation of the time series, and $\langle {\rm SFR} \rangle$ is the average SFR of the simulation. The averaging over short timescales (5 Myr, 10 Myr, 25 Myr) ensure that rapid random fluctuations in the SFR, which do not affect large-scale movement of gas due to feedback, are not counted as additional burstiness.

In Fig.~\ref{fig:burstysfr}, we show the dependence of the burstiness on the anisotropy factor $\gamma$ for various averaging durations $\tau_{\rm avg}$ = 5 Myr, 10 Myr, and 25 Myr. Using this metric, we find that for any choice of $\tau_{\rm avg}$ within this range, isotropic runs with lower $\gamma$ have bursty SF histories, and runs with $\gamma > 1$ have substantially less bursty SF histories. These findings are robust to several free parameters of the simulation including the density threshold for star formation, the star formation efficiency and the numerical resolution (see Appendix~\ref{app:numerics} for details).

We also show the dependence of the total stellar mass on $\gamma$ in the top panel of Fig.~\ref{fig:burstysfr} and find that a preferentially off-plane momentum injection (higher $\gamma$) is associated with a lower stellar mass. The effect is noticeable, with the final stellar mass differing by roughly half an order of magnitude between the highest and lowest anisotropy factors considered. From this, we conclude that out-of-disk momentum injection is associated with a less bursty and lower overall rate of star formation. 

\subsection{Gas Morphology and Bulk Motions}

The increased burstiness seen in runs with isotropic or in-disk favored momentum injection can be explained by differences in how bulk gas flows within the gas disk are induced by the SN momentum injection. We consider the flows generated parallel to the disk, or in the radial direction, defined by
\begin{equation}\label{eq:mdotR}
    \Dot{M}_{R} = \frac{1}{L} \int {\rm d}m \, {\bf v} \cdot {\bf \hat{R}} \ ,
\end{equation}
where $\hat{R}$ refers to the unit vector of the 2D cylindrical radial coordinate, and flows generated perpendicular to the disk, or in the vertical direction, defined by
\begin{equation}\label{eq:mdotZ}
    \Dot{M}_{z} = \frac{1}{L} \int {\rm d}m \, {\bf v} \cdot \pm {\bf \hat{z}} \ ,
\end{equation}
Here ${\bf v}$ is the velocity of the gas, and the integrals are taken over thin shells of thickness $L = 50\, {\rm pc}$, centered about the surface of a cylinder with diameter $2$ kpc (radius $1$ kpc) and height $0.5$ kpc ($0.25$ kpc above and below the center), centered about the center of mass of all star particles. In Fig.~\ref{fig:gasmovement}, we show a diagram of the cylinder of interest (left), and the mass flux rates of gas across the shell (right). For the in-disk favored (\textit{zeta0.1}, top right), isotropic (\textit{zeta1}, middle right) and out-of-disk favored (\textit{zeta10}, bottom right) runs, the radial (parallel to disk plane) gas mass flux rates are shown in blue, and the vertical (perpendicular to disk plane) gas mass flux rates are shown in green. To study the way star formation episodes affect these bulk flows, these mass flux rates are compared to the SFR in the cylinder's interior, shown in red.

\begin{figure*}
\centering
\includegraphics[width=0.95\textwidth]{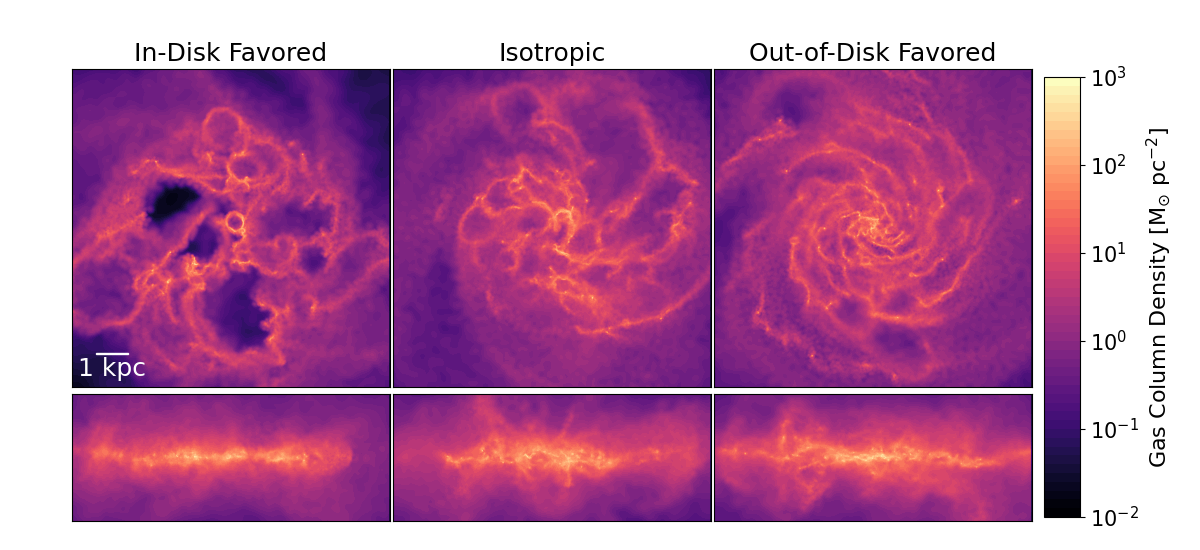}
    \caption{Gas column density of the in-disk favored (left), isotropic (center), and out-of-disk favored (right) runs at snapshot time 1.2 Gyr. The box width of each panel is 10 kpc on a side. Runs with higher injected momentum within the disk exhibit more diffuse structures and larger gas void regions, but slightly less turbulence in the vertical directions, than runs with more injected momentum out of the disk.}
    \label{fig:gasmap}
\end{figure*}

Since the gaseous disk extends to a diameter of roughly 8 kpc, the cylindrical region in question is deeply embedded inside the disk. Thus, the rates calculated in Eqs. \ref{eq:mdotR} and \ref{eq:mdotZ} trace large-scale flows \textit{within} the disk, rather than outflows. Typically, the cylinder roughly encloses $50 \, \%$ of the stellar mass and $5$ to $20 \, \%$ of the baryonic mass of the galaxy, with lower baryonic fractions occurring during intense outflow episodes, where the gas is temporarily expelled from the central regions.

In the top and middle panel, we find that in the in-disk favored and isotropic run, a significant proportion of momentum is injected within the plane of the disk, leading to large amounts of gas being moved in bulk flows parallel to the disk plane. These bulk gas flows inside the disk can also drive additional star formation bursts due to gas compression, leading to additional stars formed in these runs, and drive instability within the disk, perpetuating the cycle of radially outward gas motions and bursty star formation. Note the clear correlation between inward gas motions (dips in the blue and green curves in Fig.~\ref{fig:gasmovement}), star-formation activity (peaks in the red curve), and the subsequent outward gas motion (peaks in the blue curve). The timing of the starbursts is consistent with those predicted in \citet{Cenci23Compaction}, which states that a centrally concentrated gas distribution in a dwarf is a good indicators that a starburst will occur. Outflow episodes occur quite rapidly, with the blue and green peaks lasting $\sim$25 Myr (compared to an orbital timescale of $\sim$125 Myr at the cylinder's radius).

Conversely, in the out-of-disk favored run (bottom panel), the momentum is primarily injected outside the disk, so that gas compression induced by bulk radial flows are less prevalent, along with a less bursty star formation history. This indicates that radial, not vertical, gas flows are primarily responsible for driving additional star formation and turbulence. Interestingly, even though the SNe do not lead to bursty SFR or large-scale gas flows in the out-of-disk favored run, the stellar feedback is still capable of regulating SFR to similar or lower levels than the isotropic or in-disk injection runs. Thus, neither bursty star formation histories, nor large-scale outflows, are \textit{required} to suppress star formation (e.g., by clearing out all the gas within a local region). Instead, it is enough to have a low, steady level of SFR that prevents the runaway formation of other star-forming clumps. In our model, photoionization (which is modelled in the same way regardless of the preferential direction of the SN feedback) is also partially responsible for star-formation regulation in the scale of dwarfs \citep[e.g., see also ][]{SmithSijackiShen18}.

\begin{figure*}
\centering
\includegraphics[width=0.95\textwidth]{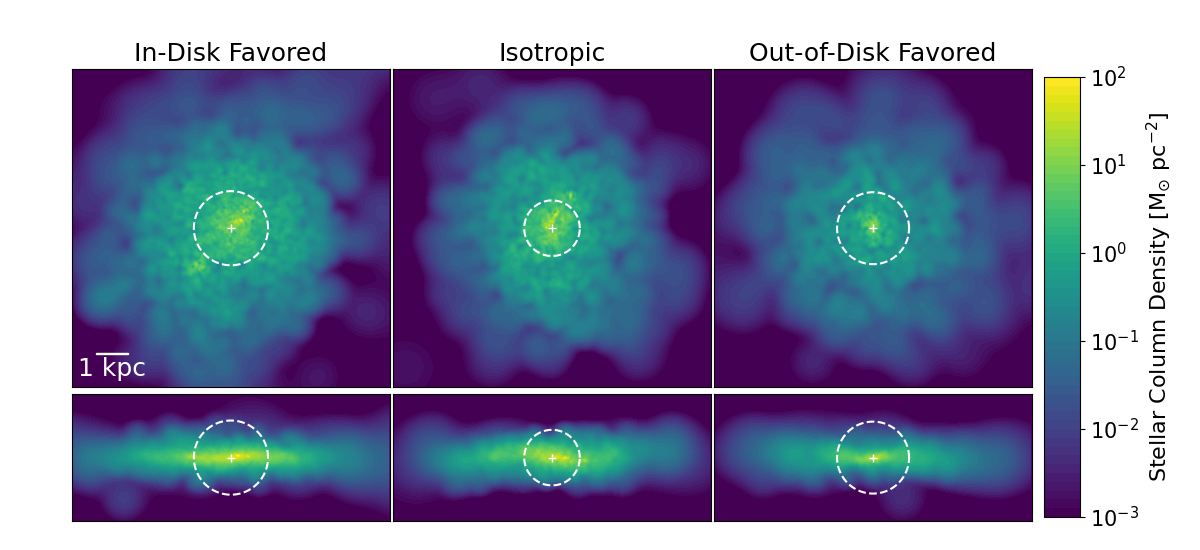}
    \caption{Stellar column density} of the in-disk favored (left), isotropic (center), and out-of-disk favored (right) runs for snapshot at 1.2 Gyr. The stellar half-mass radius about the stellar center of mass is shown by the white dashed lines. The box width of each panel is 10 kpc on a side.
    \label{fig:starmap}
\end{figure*}

We may also quantify the efficiency of the stars at generating bulk gas flows via a mass loading factor $\eta$, defined as the ratio of the total gas mass expelled by SFR-driven flows, to the total stellar mass formed. For a given run, we calculate
\begin{equation}
    \eta = \frac{1}{M_{*}} \int \, dt \, \left( \Dot{M}_{R,{\rm Out}} + \Dot{M}_{z,{\rm Out}} \right) \, ,
\end{equation}
where $M_{*}$ is the total stellar mass formed within the cylinder, and $\Dot{M}_{R,{\rm Out}}$ and $\Dot{M}_{z,{\rm Out}}$ are the quantities in Eqs. \ref{eq:mdotR} and \ref{eq:mdotZ} integrated only over outward-moving particles within the shell. The time integral is done over the total simulation time of 2 Gyr. Calculating this value\footnote{Note that this value should not be compared to typical measurements of the mass loading factors in the literature since we calculate $\eta$ for a cylinder deeply embedded in the gas disk.} for the three runs pictured in Fig. \ref{fig:gasmovement} yields $\eta \sim 40$ for the in-disk favored run, $\eta \sim 60$ for the isotropic run, and $\eta \sim 80$ for the out-of-disk favored run. This indicates that individual stars in the out-of-disk favored run are more \textit{efficient} at pushing and driving gas flows than the in-disk favored and isotropic runs. Conversely, isotropic or in-disk injection models result in increased turbulence within the disk which is further exacerbated by the momentum resulting from a larger number of stars formed than in the out-of-disk model.

The effect of the bulk gas motion cycles is further supported by the resulting structure of the gas disk, shown in Fig. \ref{fig:gasmap}. Significant differences in the structure can be seen as a result of changing the anisotropy factor. In particular, in-disk (\textit{zeta0.1}, left panel) and isotropic (\textit{zeta1}, middle panel) momentum injection exhibit a turbulent/disrupted gas morphology, including large regions or bubbles devoid of gas with radius close to $\sim 1$ kpc in the most extreme cases. This kind of structure is not present in the case of out-of-disk favored injection (\textit{zeta10}, right panel), which shows a smoother and more regular gas disk without the presence of large gas bubbles.

\subsection{ Stellar Morphology and Kinematics }

The effect of different schemes for SNe feedback injection extends also to noticeable changes in the stellar morphology of the simulated dwarf. Previous claims in the literature \citep[e.g.,][]{Stinson07,ElBadry16} have highlighted that bursty star formation histories tend to be associated with a more diffuse distribution of stars, perhaps also contributing to the erasing of age and metallicity gradients in simulated dwarf galaxies \citep[e.g. ][]{Mercado2021}

\begin{figure}[h]
\includegraphics[width=0.95\columnwidth]{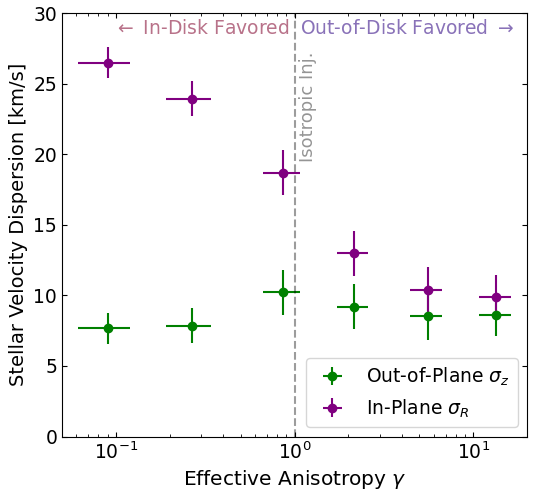}
    \caption{Vertical (green) and in-plane (purple) stellar velocity dispersion of as a function of anisotropy. Changing the degree of anisotropy does not affect the vertical velocity dispersion of stars, even when we inject more momentum in that direction. Conversely, the radial velocity dispersion of stars is inversely correlated with the anisotropy factor, so that higher radial velocity dispersions are observed when momentum is injected preferentially within the disk. This may occur either as stars tend to be born on more radial trajectories, or as stellar orbits are dynamically heated by radial flows.}
    \label{fig:vdisp}
\end{figure}

To quantify the stellar structure in our simulations, we start by showing projections of the stellar morphology of the in-disk favored (left), isotropic (middle), and out-of-disk favored (right) runs in Fig.~\ref{fig:starmap}. As a crude measure of how centrally concentrated the stars are in the dwarf, we calculate the stellar half-mass radius $r_{\rm half}$ as the spherical radius from the stellar center-of-mass in which half of the stellar mass of the galaxy is contained. We show $r_{\rm half}$ as a white dotted circle in each of the panels in Fig. \ref{fig:starmap}. $r_{\rm half}$ does not differ significantly between runs (we have explicitly checked this as a function of time) and, in general, the projected stellar structure does not have any obvious indication or trend with the momentum injection model.

However, Fig.~\ref{fig:vdisp} shows that the kinematics and orbital properties of the stars have an imprint of the chosen feedback injection scheme. We plot, for each simulated galaxy, its velocity dispersion in the 2D radial direction (purple) and the $z$-direction (green), as a function of its anisotropy factor $\gamma$, where the mean value and uncertainty are calculated over all snapshots in the latter 1.5 Gyr of evolution. Preferential injection of momentum in the plane of the disk (lower-$\gamma$ runs) results in noticeably higher stellar {\it radial} velocity dispersion, presumably as a combination of two effects: new stars being born from gas with already large radial motions, and the input of dynamical heating to the population of already formed stars as a result of radial flows and compression waves.

Interestingly, the fact that the stellar velocity dispersion in the $z$-direction shows little to no dependency on $\gamma$ (green symbols) suggests that stellar kinematics do not always trace the properties of outflowing gas (although they do trace the kinematics of star-forming gas). If, contrarily, newly formed stars inherited the properties of the bulk gas flow, then the degree of SN anisotropy should affect the stellar kinematics, in the same way that it affects the vertical outflow rates shown in Fig.~\ref{fig:gasmovement}. However, this is not the case, as the stellar $\sigma_z$ appears to be independent of the SN anisotropy; thus, the stellar motions and the total gas motions are not well-coupled to each other. 

\subsection{Distribution of Dark Matter}

The difference in the amount and morphology of the bulk gas motions discussed in the previous sections has the potential to impact not only the baryonic morphology of the dwarf, but also its dark matter distribution. Bursty star formation histories that are associated with locally-dominant gas expulsion can lead to the transformation of cuspy dark matter halos into lower density dark matter cores \citep[e.g.,][]{PontzenGovernato12,BenitezLlambay19}. However, models resulting in a smoother star formation history (even when considering the exact same baryonic physics) could in principle preserve the characteristic $\lcdm$ cuspy profiles, changing the theoretical expectations for dwarf galaxies. 

\begin{figure}[h]
\includegraphics[width=0.99\columnwidth]{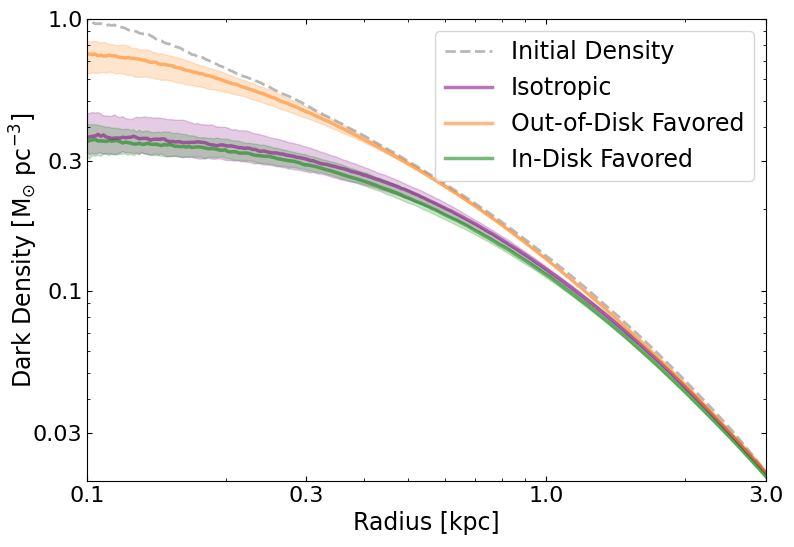}
    \caption{Dark matter density profiles between 1 and 2 Gyr. The median density across snapshots at each radius is plotted as a solid line, and the interquartile range at each snapshot is shaded. Consistent with the understanding of core formation, the out-of-disk favored run, which has a non-bursty SF history, has a dark matter cusp, and the isotropic and in-disk favored runs, which have bursty SF histories, have dark matter cores. For comparison, the initial cuspy density profile (measured as the median density from the stellar center-of-mass over the first 200 Myr, similar across all runs considered) is shown by the grey dashed line.}
    \label{fig:densprof}
\end{figure}

To address this, we compare in Fig. \ref{fig:densprof} the dark matter density profile of our simulated dwarfs in the in-disk favored (\textit{zeta0.1}, green), isotropic (\textit{zeta1}, red) and out-of-disk favored (\textit{zeta10}, orange) runs. These profiles have been calculated as the average dark matter density in the last $1$ Gyr to average out temporal fluctuations. Thick lines indicate the median while the shaded regions highlight the interquartile range. 
We find that lower-$\gamma$ runs, which are associated with more bursty SFR and in-plane gas flows, tend to have dark matter cores compared to higher-$\gamma$ runs, for which less bursty SFRs were measured (see Fig.~\ref{fig:burstysfr}) and that retain a substantially more cuspy profile. 
For reference, the dashed gray line indicates the initial profile which corresponds to a cuspy NFW profile with concentration $c=15$. This correlation between the mode of star formation and the inner dark matter distribution is consistent with the expectation from the literature, as discussed above.

Notably, the distinction in the cuspy vs.~cored dark matter prediction holds when considering exactly the same {\it physics} of the ISM. While previous work has already highlighted that some baryonic treatments would result in cores while others would produce cusps \citep[see e.g.,][]{Bose19} over similar mass ranges, the treatment of the baryons from the models considered was different, opening the possibility that the core formation was purely the result of including more detailed or more ``realistic" star-formation and feedback models. This degeneracy is underscored by models resolving SN directly but not producing cores \citep{CosmoLyra2022}. In short, our results confirm that not only the physics and treatment for the ISM play a role, but even the detailed numerical choices for, e.g., distributing the momentum of the stellar feedback may cause profound changes in the predictions for the density in the inner regions of dwarfs. 

This also suggests that implementations of ``universal" corrections for modifying the cuspy NFW profile into a lower density core that depend solely on the stellar mass $M_*$ or stellar-to-halo mass ratio $M_* / M_{\rm vir}$ \citep[as proposed, e.g., by][]{DiCintio14} might describe the behavior for a given feedback model or family of baryonic models very well, but are not necessarily applicable to others. For example, all runs considered in Fig. ~\ref{fig:densprof} have a stellar-to-halo mass ratio of $\sim 10^{-2.5}$, a regime where the inner region is predicted to be cored (with a profile slope of $-0.3$ to $-0.1$). This is consistent with the isotropic and out-of-disk favored runs at radii below about $\sim 500$ pc, but the profile slope of the out-of-disk favored run at this distance is nearly $-1$, consistent with a cusp instead. We hasten to add, however, that this implication regarding universal modifications to the dark matter profile still needs further confirmation from runs within the cosmological set-up.

\section{Robustness of the results to changes in the cell weighting scheme}\label{sec:weighting}

As stated in Section~\ref{sec:simulations}, it is also possible to change the magnitude of the momentum distributed across gas cells by changing the weighting function, as described by the weighting factors $w_i$. The default cell weighting scheme in {\small SMUGGLE} is based on the solid angle sub-tended by the gas particles from the target star, in a way that nearby cells or cells with a large area have larger weights \citep[see Sec. 2 in][]{SMUGGLE}. This scheme follows other codes that include a similar weighting, like {\small FIRE-2} \citep[][]{FIRE2}. 

However, other choices of the cell weighting scheme are possible, including mass-based and volume-based weighting. In a hydrodynamic solver with roughly equal-mass gas cells, these physically correspond to favoring injection towards dense gas regions, and towards diffuse gas regions, respectively \citep[see discussion in][]{FIRE2}. As the energy-conserving phases of SNe are not explicitly resolved in simulations at our resolution, and because local environments in which SNe go off are potentially inhomogeneous, it is not immediately clear what gas the feedback momentum will most effectively couple to. As such, the choice of cell weighting scheme should be understood as an \textit{assumption} made in the subgrid model. In this section, we explore how these three choices (solid angle, volume, and mass) of cell weighting scheme influences the feedback results. The formulae for these weighting schemes are described in Section~\ref{sec:simulations} (in particular, see Eq.~\ref{eq:omega} and the subsequent discussion). 

Nominally, the momentum injected by a supernova should be the same across all SN events. However, as discussed in section \ref{subsec:snfeed}, a number of 'numerical post-processing' procedures must be applied when imparting the momentum to a neighboring gas cell, to correct for the unresolvedness of the energy-conserving phase. In particular, the momentum going into a gas cell is boosted by the factor $\beta$ \citep[as per Eq. \ref{eq:boostfac}, following][]{Hopkins18SNFeed}, and furthermore, momentum is not added if the gas cell is more than $r_{\rm SB} = 1024 \, {\rm pc}$ from the star particle. As such, the true amount of momentum injected by star particles depends on the configuration of individual gas cells at the time of the SN event, which can vary across different runs.

\begin{figure}[h]
\includegraphics[width=0.95\columnwidth]{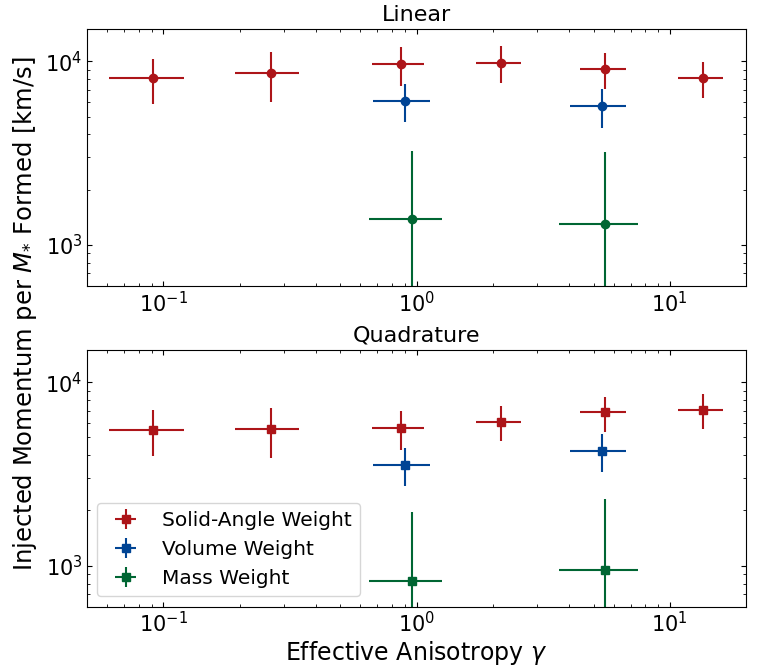}
    \caption{Magnitude of momentum injected by SNe per mass of stars formed, for each run over 2 Gyr. We calculate the total magnitude in two ways: linear (top panel, Eq. \ref{eq:ptotlin}), and in quadrature (bottom panel, Eq. \ref{eq:ptotquad}). For the same weighting scheme, there is no dependence of the magnitude of momentum injected on the anisotropy factor $\gamma$. However, between solid-angle weighting (red), volume weighting (blue), and mass weighting (green), the magnitude of momentum injected can differ by as much as a factor of $\sim$5. These behaviors hold using both calculation methods (although the values do depend on the calculation method).}
    \label{fig:totmom}
\end{figure}

In fact, we find that, purely due to these numerical post-processing effects the momentum typically injected by a single star particle {\it may significantly differ} across runs. In particular, the momentum injected by SNe, per stellar mass formed, can vary by almost an order of magnitude when different cell weighting schemes are considered, independently of the anisotropy factor $\gamma$ in that run. In Fig. \ref{fig:totmom}, we plot for each simulation, the mean and standard deviation over all star particles of the quantity
\begin{equation}
    {\rm Injected \, Momentum \, per} \, M_{*} =
    \left(
    \frac{p_{\rm tot}}{M_{\rm init}}
    \right)_{\rm stars},
\end{equation}
where $p_{\rm tot}$ is the total magnitude of momentum injected by a given star particle, and $M_{\rm init}$ is its initial mass.

\begin{figure*}[t]
\centering
\includegraphics[width=0.95\textwidth]{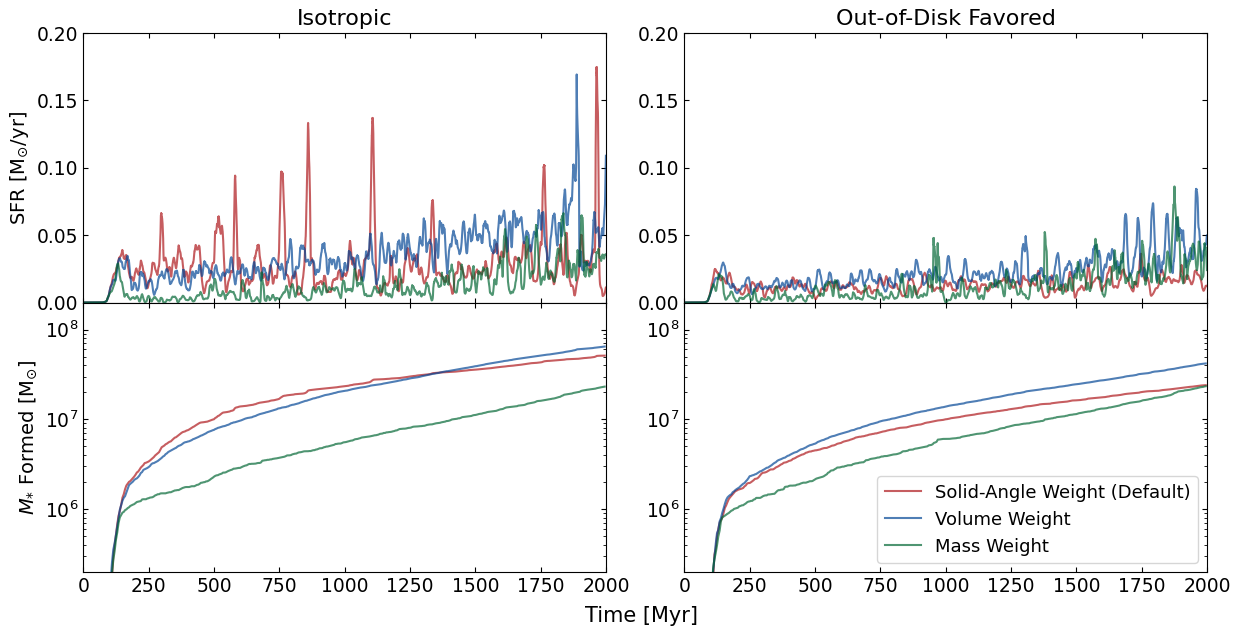}
    \caption{SFR histories (top) and stellar mass formed (bottom) for runs with different weighting schemes: solid angle (red), volume (blue) and mass (green). The left and right panels correspond to isotropic momentum injection ($\gamma \approx 1$) and out-of-disk favored momentum injection ($\gamma > 1$). In general isotropic injection results on more bursty SFR histories than out-of-disk favored ones. Solid-angle based criteria assuming isotropic momentum injection maximizes the burstiness of star-formation. The total stellar mass formed can vary by factors $2$-$3$ between runs with different numerical choices.}
    \label{fig:WeightedSFH}
\end{figure*}

Since the injected momentum is a vector quantity, there is no canonical way of calculating the its total magnitude. Thus, we define two possible ways of quantifying $p_{\rm tot}$: using the linear method
\begin{equation}\label{eq:ptotlin}
    p_{\rm tot,Lin.} =
    \sum_{d} \sum_{\rm SNe} |\delta p_{d}|
\end{equation}
\noindent
and the quadrature method
\begin{equation}\label{eq:ptotquad}
    p_{\rm tot,Quad.} =
    \sqrt{
    \sum_{d}
    \left(\sum_{\rm SNe} \, |\delta p_{d}| \right)^2
    },
\end{equation}
where $\delta p_d$ for $d=x,y,z$ are the momenta injected during a given supernova event in the $x$-, $y$-, and $z$-directions, respectively. Note that this calculation reflects the mean and dispersion {\it normalized by stellar mass formed}, so the comparison is meaningful even if the runs formed different amounts of stellar mass by the end. For our default choice, opening-angle weight (also referred to as ``omega"-weight, red symbols), we show that the median input as a function of anisotropy (effective anisotropy, $\gamma$) remains quite constant, suggesting that the typical momentum input for this weighting scheme is independent of the directionality ($\gamma$) chosen. The trend remains in both, the linear (top panel) or in-quadrature (bottom) measurement of momentum.

We also compare to a volume weighting scheme (blue) and mass weighting scheme (green) for the isotropic and out-of-disk momentum injection ($\gamma \sim 1$ and $10$, respectively). We have explicitly checked that the effective anisotropy of momentum injection is comparable across all weighting schemes (see Table~\ref{tab:allsims}), which then allows for a fair comparison between total momentum injection at a given $\gamma$. Fig.~\ref{fig:totmom} shows a systematic shift of the momentum effectively imparted per star depending on the choice of weighting function, with a volume-weighting criteria returning values slightly below but comparable to the omega weighting scheme, but a mass weighted injection results in up to $5$ times less momentum input per star. The effect also seems independent of $\gamma$, at least for the two explored values here. It is worth noting, that of all weighting methods, the opening angle criteria adopted by default in {\small SMUGGLE} seems to maximize the momentum injection into the ISM. Also important is the observation that the choice of cell-weighting may be well physically motivated, but can result in unwanted pathological behaviors. For instance, mass-weighting schemes in poorly-resolved disks may lead to the creation of artificially large bubbles \citep{Torrey2017}, perhaps making such a choice unpreferred for runs with intermediate to low resolution. 

These differences in momentum input have a noticeable effect on the star formation histories of our simulated dwarfs. We show the evolution of the SF history (top) and the total stellar mass (bottom) in Fig.~\ref{fig:WeightedSFH} for these three weighting schemes, and choosing isotropic momentum injection (left) and out-of-disk injection (right). The {\small SMUGGLE} default choice, omega-weighted and isotropic momentum injection, leads to the most bursty star formation of all runs (red curve, top left panel). 

All weighting schemes seem to self-regulate star formation, at least for the $2$ Gyr explored here, with the volume and mass weighting models predicting a slight but steady increase of SFR with time. Differences in weighting and directionality of momentum injection can cause variations by factors $2$-$3$ of the total amount of stars formed, with isotropic injection resulting in systematically larger stellar masses for all weighting schemes, while  at fixed momentum injection directionality, the mass-weighting scheme results in lower $M_*$. 

Interestingly, neither the total rate of star formation nor the stellar mass are correlated with the magnitude of momentum injected shown in Fig.~\ref{fig:totmom}. For example, the mass-weighted runs have the least momentum injected per star, but also the lowest rate of star formation, contrary to expectations. Furthermore, the volume-weighted runs form more stars than the omega-weighted runs, but have less momentum injected per star particle than the omega-weighted runs. This shows that the effect of momentum injections on the star formation history is not immediately obvious; that is, it is not true that the input of more feedback momentum translates into more effective suppression of star formation (i.e., a lower stellar mass), as may be naively expected.

\section{Conclusions}\label{sec:conclusion}

Undoubtedly, the assumption of isotropy for the momentum deposition resulting from a single supernova event in a homogeneous and constant density medium is well justified. However, the ISM of galaxies is far from this idealized constant density assumption, and stellar clustering may cause the overlap and fusion of multiple bubbles leading to a complex system where the assumed isotropy for energy and momentum deposition no longer holds. The scales over which these mechanisms occur are currently under-resolved in most galaxy-scale simulations. Here we explored the systematic effects of lifting the hypothesis of isotropic bubble expansion using idealized numerical simulations of an isolated dwarf galaxy run with the {\small SMUGGLE} model. 

We implemented several different schemes to distribute and couple the momentum resulting from supernova feedback onto neighboring gas cells. More specifically, we explored independently the effects of $i)$ changing the directionality of the momentum injection (lifting the isotropic assumption) and $ii)$ studying the effect of different weighting schemes (independent of direction). We label our runs according to the excess of momentum deposited perpendicular to the disk using the parameter $\gamma$. Isotropic momentum injection corresponds to $\gamma \approx 1$, while $\gamma > 1$ and $\gamma < 1$ indicate that the momentum injection is favored perpendicular (vertical) and parallel (horizontal) to the plane of the disk, respectively.

We can summarize our most important results as follows:
\begin{itemize}
    \item The degree of anisotropy assumed in SNe momentum injection greatly affects the burstiness of the star formation history. When momentum is primarily directed perpendicular to the plane of the disk, bursty star formation is suppressed, and less total stars are formed, compared to runs in which momentum is primarily injected isotropically or within the plane of the disk.
    
    \item Bursty star formation results from runs where momentum is coupled effectively within the plane of the disk, generating in-plane bulk gas motions that create more turbulence in the disk. The gas structure tends to be more disrupted when momentum is primarily injected within the disk plane, and less disrupted when momentum is primarily injected perpendicular to the disk plane. 
    
    \item The anisotropy of the momentum injection also clearly affects the resulting stellar morphology of the dwarf. This can be observed in the 2D radial velocity dispersion of stellar particles being more than double in cases of in-disk favored and isotropic injection compared to an out-of-disk favored momentum injection scheme. Stellar orbits become clearly less circular for models where momentum is primarily coupled in the plane of the disk ($\gamma \leq 1$ models), an effect that we may attribute to a combination of higher radial velocities in star-forming gas, and increased gravitational heating from in-plane gas flows in these runs compared to higher-$\gamma$ models. 
    
    \item Runs with bursty star formation form dark matter cores, and runs with non-bursty star formation retain their dark matter cusps.  This is consistent with the existing understanding of the relationship between significant gas flows (typically associated to bursty star formation) and dark matter core formation. However, our results also suggest that in "mid-resolution" simulations that rely on subgrid SNe feedback prescriptions, the formation of dark matter cores may still depend on numerical choices. Thus, the core properties such as radial extension and inner density profile are not yet firmly established as a fundamental prediction of the physics included in a given simulation.
    
    \item Different cell weighting schemes in SNe may, due purely to numerical effects, actually lead to different amounts of momentum being injected. This qualitatively changes the star formation history and may vary the total amount of stellar mass formed by factors of $2$-$3$ within 2 Gyr of evolution. A weighting scheme based on opening angle, such as that used in the default version of {\small SMUGGLE}, seems to maximize the momentum input per stellar particle, being $\sim 5$ times larger than, for example, a scheme based on cell mass alone. As such, the cell weighting scheme must also be understood as a numerical choice (although it may still be a physically motivated choice).
\end{itemize}

Our results imply that numerical choices implicit to the sub-grid modeling of SNe feedback can play a significant role in the predicted properties of simulated dwarf galaxies. In particular, for the {\small SMUGGLE} code, we demonstrate that the level of burstiness in the star formation history is particularly sensitive to the numerical implementation of the underlying feedback physics. This has consequences for the predicted stellar and gas morphology, with more bursty runs leading to less rotational support of the gas and stars, along with a more distorted ISM characterized by large bubbles and in-disk gas flows. While the specific effects of the weighting scheme will likely vary from model to model (and depend on details of hydrodynamics solver, numerical implementation, etc), our study is an important reminder that results from numerical simulations should be scrutinized not only in light of the physics included, but also of the numerical choices made, a subject that receives considerably less attention in the literature. 

We stress that, though the parameters studied in this work are numerical, this does not mean that the numerical choices made have no physical correspondence. Indeed, the anisotropy parameter and the cell weighting scheme may be physically motivated choices, that depend on the types of gas that SNe explosions are expected to couple to most effectively. In reality, the true nature of how SNe momentum couples to ISM gas is not yet well understood, and subgrid models must accurately mimic this coupling in order to produce true predictions from simulations. Thus, in future work it may be instructive to compare these simulations with models that resolve SN explosions explicitly, such as {\small LYRA} \citep{Lyra2021,SampleLyra2022} and {\small RIGEL} \citep[e.g.,][]{Deng24RIGEL}. In such ultra-high resolution simulations, details of the momentum distribution, including the degree of anisotropy, are expected to be produced self-consistently by the resolved thermal expansion. Detailed investigations of the bubble morphology in galaxy simulations with resolved SNe might be able to better inform subgrid models that rely on numerical parameters. These parameters may depend on more detailed properties of the ISM such as directional variations in density, in addition to commonly considered "zeroth-order" parameters such as the energy released or average local density.

A possible interpretation of the results presented here is that the level of burstiness in simulated dwarfs could be over-represented at least in some galaxy formation models. For example, the lack of formation of dense and compact dwarf galaxies in most codes available today supports the scenario where more violent star-formation and feedback cycles are present in simulated dwarfs compared to the real Universe \citep[see discussion in][]{Sales22}. Similarly, as reported by \citet{ElBadry2018}, simulated dwarfs in {\small FIRE} on the scale of $M_* \sim 10^8$ - $10^9\; \rm M_\odot$ seem less rotationally supported than suggested by HI observations, also consistent with the possibility of too bursty star formation rates. In this regard, studies trying to compare the star formation cycles in observation and simulations may prove one of the most important ways to inform dwarf galaxy formation models in simulations \citep[e.g., ][]{Patel2018, Emami2019, Emami2021, Mehta2023}.

Constraining the star formation cycles is especially important not only for our understanding of dwarf galaxy morphology, but also for the underlying link between bursty star formation histories, gas flows generated and their impact on the innermost distribution of dark matter in dwarfs. In our runs with less bursty SFR, the formation of dark matter cores at the center of dwarfs was strongly suppressed, meaning that our theoretical expectations of a cusp or a core at the center of dwarfs depends at least on both, the physics included in the simulation as well as on the numerical choices made in coarsely implementing this physics in each code.
\\\\

\section*{Acknowledgements}

The authors would like to thank Phil Hopkins and Julio Navarro for insightful discussions that helped improve the first version of this draft. EZ and LVS acknowledge the financial support received through NSF-CAREER-1945310 and NSF-AST-2107993 grants. FM acknowledges funding by the European Union - NextGenerationEU, in the framework of the HPC project – “National Centre for HPC, Big Data and Quantum Computing” (PNRR - M4C2 - I1.4 - CN00000013 – CUP J33C22001170001). PT acknowledges support from NSF-AST 2346977. HL is supported by the National Key R\&D Program of China No. 2023YFB3002502, the National Natural Science Foundation of China under No. 12373006, and the China Manned Space Program through its Space Application System. TAG acknowledges support by NASA through the NASA Hubble Fellowship grant $\#$HF2-51480 awarded by the Space Telescope Science Institute, which is operated by the Association of Universities for Research in Astronomy, Inc., for NASA, under contract NAS5-26555. Computations were performed using the computer clusters and data storage resources of the HPCC, which were funded by grants from NSF (MRI-2215705, MRI-1429826) and NIH (1S10OD016290-01A1). 

\appendix

\section{Robustness of Results to Other Numerical Choices}
\label{app:numerics}

The changes in the dwarf properties that result from varying the anisotropy of SNe momentum injection are robust to changes in major simulation parameters, including the efficiency and density threshold of star formation, and the mass resolution. To verify this we have run 8 additional simulations of the same dwarf, in which these simulation parameters were varied. These runs are summarized in Table~\ref{tab:robustness}.

\begin{figure*}
\centering
\includegraphics[width=0.75\textwidth]{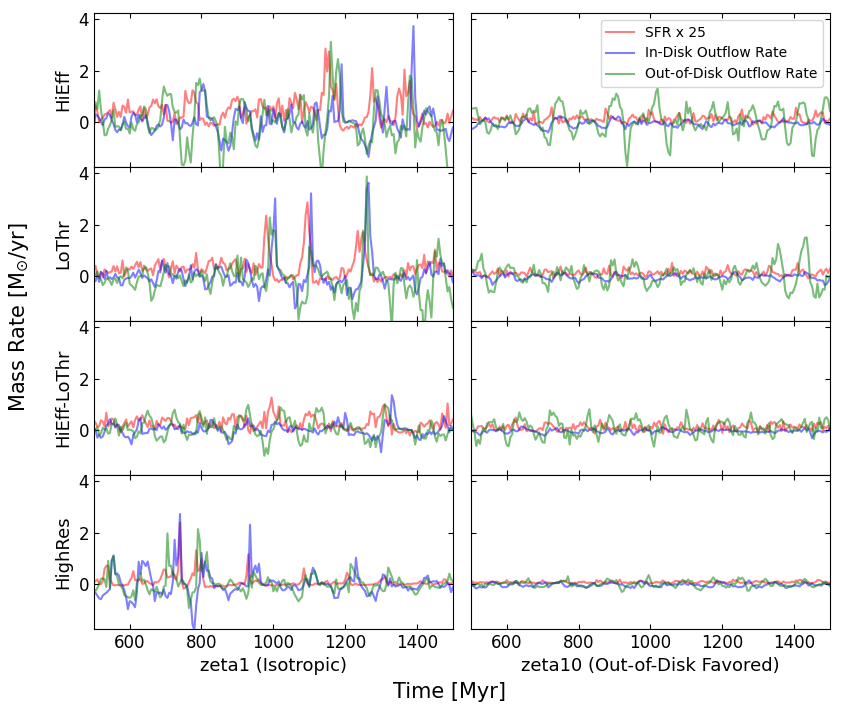}
    \caption{In-disk, or radial (blue) and out-of-disk, or vertical (green) gas flow rate and associated SFR (red) for all 8 robustness test runs, with respect to the same cylinder described in Fig.~\ref{fig:gasmovement}. In all cases, \textit{zeta1} runs have burstier SFRs and more violent gas flow patterns, whereas the \textit{zeta10} runs do not. This is the same behavior as observed in the main body of the paper, so changing the anisotropy affects the runs in this way regardless of these numerical choices explored here.}
    \label{fig:robustness}
\end{figure*}

In particular, we find that across a wide range of these parameters, varying the anisotropy factor $\gamma$ produces the same effect: in-disk favored and isotropic momentum injection produce a bursty SFR history with periods of major in-plane gas motion, whereas out-of-disk favored injection produces a much smoother SFR level with a less turbulent bulk gas flows. We show this behavior in Fig.~\ref{fig:robustness}.

\begin{table*}
\centering
\begin{tabular}{lcccr}
    \hline
    Simulation Name & Nominal SNe Anisotropy $\zeta$ & Max. SFR Efficiency & SFR Threshold [$\rm{m_H} / \rm{cm}^{3}$] & Target Gas Mass [$\msun$] \\
    \hline
    \textit{zeta1-HiEff} & 1.0 & 0.9 & 100 & 6000 \\
    \textit{zeta10-HiEff} & 10.0 & 0.9 & 100 & 6000 \\
    \textit{zeta1-LoThr} & 1.0 & 0.01 & 1 & 6000 \\
    \textit{zeta10-LoThr} & 10.0 & 0.01 & 1 & 6000 \\
    \textit{zeta1-HiEff-LoThr} & 1.0 & 0.9 & 1 & 6000 \\
    \textit{zeta10-HiEff-LoThr} & 10.0 & 0.9 & 1 & 6000 \\
    \textit{zeta1-HighRes} & 1.0 & 0.01 & 100 & 500 \\
    \textit{zeta10-HighRes} & 10.0 & 0.01 & 100 & 500 \\
    \hline
\end{tabular}
\caption{List of simulations used in our robustness analysis, and the parameters used in each.}
\label{tab:robustness}
\end{table*}

\bibliographystyle{aasjournal}
\bibliography{biblio}

\label{lastpage}

\end{document}